\documentclass[pdflatex,iicol,sn-mathphys-ay]{sn-jnl}

\let\orcidlogo\relax
\usepackage{orcidlink}
\usepackage{amsmath,amssymb,amsthm,mathtools}
\usepackage{graphicx,booktabs}
\usepackage{algorithm}
\usepackage{algorithmicx}
\usepackage{algpseudocode}

\newcommand\given[1][]{\:#1\vert\:}

\newcommand{\Var}{\operatorname{Var}}
\newcommand{\AND}{\mathbin{\mathrm{and}}}
\newcommand{\OR}{\mathbin{\mathrm{or}}}

\begin{document}

\title[Hash-augmented adaptive multilevel splitting Monte Carlo]{Hash-augmented adaptive multilevel splitting Monte Carlo algorithm for accurate estimation of two-sample permutation test \emph{p}-values}

\author[1]{\fnm{Nikita} \sur{Golikov}}
\author[2,1]{\fnm{Vladimir} \sur{Sukhov}~\orcidlink{0000-0002-5169-1433}}
\author[3]{\fnm{Gennady} \sur{Korotkevich}}
\author*[2]{\fnm{Alexey} \sur{Sergushichev}~\orcidlink{0000-0003-1159-7220}}\email{alsergbox@gmail.com}

\affil[1]{\orgname{ITMO University}, \orgaddress{\country{Russia}}}
\affil[2]{\orgdiv{Department of Pathology and Immunology}, \orgname{Washington University in St. Louis}, \orgaddress{\country{USA}}}
\affil[3]{\orgname{Independent researcher}, \orgaddress{\country{USA}}}

\abstract{
Nonparametric permutation tests are widely used for statistical analysis. However, exact computation of test \emph{p}-values can be algorithmically challenging, particularly for custom tests with complex test statistics. In contrast, Monte Carlo sampling can be easily applied to any test statistic, but it suffers from poor relative accuracy when estimating small \emph{p}-values, interfering with multiple hypothesis testing correction and leading to other issues. In this work, we present a hash-augmented adaptive multilevel splitting Monte Carlo algorithm that enables accurate estimation of arbitrarily small \emph{p}-values in two-sample permutation tests. Using the Kolmogorov--Smirnov and the Mann--Whitney U tests as examples, we highlight potential pitfalls related to the discreteness of the test statistic distribution and show how to address them. By comparing with an exact algorithm, we demonstrate the accuracy of the \emph{p}-value estimates provided by the proposed algorithm and the validity of the associated confidence intervals.
We provide a reference implementation of the proposed algorithm in the 
Python package \texttt{hamstest}, which allows \emph{p}-value estimation for a user-defined statistic.
}


\maketitle

\section{Introduction}

Nonparametric tests are widely used for testing statistical hypotheses, as they make minimal assumptions about the underlying data distribution. 
Common examples include Fisher’s exact test, the Mann--Whitney U test, and the Kolmogorov--Smirnov test, which are frequently applied to compare two groups of observations (e.g., test and control groups).
Whereas Fisher’s exact test has a simple and efficient procedure for calculating exact \emph{p}-values, the others require more sophisticated algorithms, so many workflows 
rely on asymptotic approximations or Monte Carlo sampling to estimate \emph{p}-values.

Due to its flexibility, Monte Carlo sampling is a common approach for approximating \emph{p}-values in nonparametric tests with custom statistics across different areas of science: 
in genomics~\citep{subramanian_gene_2005}, neuroscience~\citep{maris_nonparametric_2007}, ecology~\citep{clarke_non-parametric_1993, anderson_new_2001}, climate science~\citep{livezey_statistical_1983}, and machine learning~\citep{yeh_more_2000,ojala_permutation_2010}.
The drawback of this approach is its poor scalability when approximating very small \emph{p}-values. For example, estimating a \emph{p}-value on the order of $10^{-10}$ with a relative accuracy of 100\% would require generating $10^{10}$ samples, which is highly impractical.
Nevertheless, accurate approximation of such small \emph{p}-values is critical when many tests are performed simultaneously: it ensures adequate statistical power after multiple hypothesis correction and allows for better prioritization of analysis results.

Approaches to estimating rare event probabilities using random sampling were first developed in the context of physics~\citep{kahn_estimation_1951,cerou_adaptive_2019}. One such technique is importance splitting, which involves dividing the sampling space into a sequence of importance levels. This idea evolved into \emph{multilevel splitting}~\citep{glasserman_multilevel_1999}, which relies on expressing the probability \( \Pr(\mathcal{A}) \) of a rare event \( \mathcal{A} \) as a product of conditional probabilities over nested events \( \mathcal{A}_i \subset \mathcal{A}_{i-1} \): 
\begin{equation}
\Pr(\mathcal A)
:= \Pr(\mathcal A_k)
= \Pr(\mathcal A_1)
  \prod_{i=2}^{k} \Pr(\mathcal A_i \mid \mathcal A_{i-1}).
\end{equation}
With a suitable choice of the intermediate events \( \mathcal{A}_i \), each conditional probability \( \Pr(\mathcal{A}_i \mid \mathcal{A}_{i-1}) \) can be effectively estimated using standard Monte Carlo methods.
\emph{Adaptive multilevel splitting}~\citep{cerou_adaptive_2007,cerou_adaptive_2019} further refines this approach by dynamically constructing the sequence of nested events during sampling so that each conditional probability \( \Pr(\mathcal{A}_i \mid \mathcal{A}_{i-1}) \) is approximately equal to a predefined constant, such as \( 1/2 \). This results in a logarithmic number of levels needed to estimate arbitrarily small probabilities.

While multilevel splitting is typically described for continuous probability spaces, this work focuses on potential pitfalls that arise when the technique is applied to discrete spaces, which are inherent to nonparametric tests. 
For simplicity, we use the two-sample Kolmogorov--Smirnov and Mann--Whitney U tests as examples.
Previously, we demonstrated the utility of adaptive multilevel splitting for efficient and accurate
calculation of \emph{p}-values in Gene Set Enrichment Analysis (GSEA), which is a generalization of the Kolmogorov--Smirnov test used for the analysis of genomics data~\citep{korotkevich_fast_2021}.
Here we show that the adaptive multilevel splitting scheme originally implemented in~\cite{korotkevich_fast_2021} can fail for certain inputs when the distribution of the statistic exhibits large jumps. 
To address this, we propose a new adaptive multilevel splitting algorithm for estimating two-sample permutation test p-values which augments the statistic with a hash value, greatly reducing the discreteness of the distribution.
In addition, we explore several design options for the algorithm and their impact on computational efficiency and approximation accuracy. Based on these results, the proposed \emph{hash-augmented adaptive multilevel splitting} algorithm avoids the identified pitfalls, accurately converges to the target values, and remains highly efficient.
The proposed algorithm is implemented as a Python package \texttt{hamstest}, which enables users
to apply the algorithm to their custom statistics, while accurately handling challenges associated
with discrete spaces.
\texttt{hamstest} is open source and freely available under the MIT license.

\section{Preliminaries}

\subsection{Problem statement}

Consider two samples \( X = (X_1, X_2, \ldots, X_n) \) and \( Y = (Y_1, Y_2, \ldots, Y_m) \), independently drawn from distributions \( F \) and \( G \), respectively, and the task of testing the null hypothesis \( H_0 : F = G \). 
To address this question, we can define a test statistic \( S(X, Y) = \gamma \in \mathbb{R} \) and ask how likely it is to obtain a value of the test statistic at least as extreme as~\( \gamma \). For simplicity, we consider the one-sided case, that is, calculating the \emph{p}-value as the probability \( \Pr(S \ge \gamma) \). We also assume $n \le m$.

The permutation test approach defines the above probability by considering 
all possible permutations of the input data labels.
However, the test statistic \( S(X, Y) \) commonly does not depend on the order of elements within the samples \( X \) and \( Y \)—the situation we will consider in this work—so combinations can be used instead of permutations. 
More formally, let \( Z = (X, Y) \) be the combined sample,
which, under the null hypothesis \( H_0 \), is drawn from a common distribution \( F = G \).
Let \( [n+m] = \{1, \ldots, n+m\} \), and define \( Z[A] \) as the subsequence of \( Z \) indexed by \( A \subset [n+m] \) in sorted order. We then define \( S(A) = S(Z[A], Z\bigl[[n+m ] \setminus A \bigr]) \).
Then the probability \( \Pr(S \ge \gamma) \) can be defined over the space \( \mathcal{P}_{=n}([n+m]) \) of all \( n \)-element combinations from~\( [n+m] \):
\begin{equation}
\label{eq_prob_comb_short}
\Pr(S \ge \gamma) = \frac{1}{\binom{n+m}{n}} \sum_{A \in \mathcal{P}_{=n}([n+m])} \left[ S(A) \ge \gamma \right] \text{.}
\end{equation}

In this work, we consider two examples. The first example is the one-sided two-sample Kolmogorov--Smirnov test with the statistic
\begin{equation}
\label{eq_def_dpos}
D^+ = \sup_x \left( F(x) - G(x) \right),
\end{equation}
where \( F \) and \( G \) are the empirical distribution functions of the samples \( X \) and \( Y \), respectively:
\begin{align}
\label{eq_def_dpos_fg}
F(x) &= \dfrac{\#\{ i \mid X_i \le x \}}{n}, &
G(x) &= \dfrac{\#\{ i \mid Y_i \le x \}}{m}.
\end{align}
Our goal is to calculate the \emph{p}-value of the one-sided two-sample Kolmogorov--Smirnov test, 
that is, the probability
\begin{equation}
\label{eq_prop_ks}
\Pr(D^+ \ge \gamma \mid H_0)
\end{equation}
for the observed value of the statistic \( \gamma = D^+(X, Y) \) based on the two samples \( X \) and \( Y \).

The second example is the Mann--Whitney U test. Define the pooled rank function $R \colon [n + m] \to \mathbb{R}$ over the combined sample $Z$ as follows: $R(i)$ is equal to the position of $Z_i$ in sorted vector $Z$. 
In the case of ties, a common approach is to set all ranks of tied elements to the average of their positions.
Now, the one-sided statistic $U_1$ is equal to $R_1 - \frac{n(n+1)}{2}$, where $R_1$ is the sum of ranks for the first sample $\sum_{i \in [n]} R(i)$. The one-sided test computes $\Pr(U_1 \ge \gamma \given H_0)$ for the observed value $U_1(X, Y) = \gamma$.

\subsection{Exact \emph{p}-value calculation}
\label{sec_exact}

In statistical hypothesis testing, exact \emph{p}-values are preferable because they eliminate approximation error, especially in the tails of the null distribution. 
For a permutation test, naïvely enumerating all \( \binom{n+m}{n} \) labelings incurs exponential time and is therefore impractical.  
However, for certain test statistics, exact and reasonably efficient algorithms are possible.

For the Kolmogorov--Smirnov statistic, Hodges’ classic dynamic programming method computes the exact null distribution in \( O(mn) \) time~\citep{hodges_significance_1958} and is implemented in SciPy, although that implementation reverts to an asymptotic approximation once intermediate values overflow floating-point precision. 
A more recent fast Fourier transform-based algorithm~\citep{dimitrova_computing_2020} achieves \( O(n^2 \log n) \) complexity, yet is bounded by machine epsilon and cannot report \emph{p}-values smaller than approximately \( 10^{-16} \).  

For the Mann--Whitney U test, the time complexity of standard algorithms is $O(n^2 m^2)$. Individual algorithms such as by~\cite{streitberg_exact_1984} or by~\cite{loffler_mannwhitneyu_1983} vary in their memory usage and support for ties.
Similarly to the Kolmogorov--Smirnov statistic, fast Fourier transform-based algorithms
are also possible, achieving the time complexity of $O(n^2 m)$, but also limited by the machine epsilon~\citep{nagarajan_reliability_2009}.

The examples of the Kolmogorov--Smirnov and Mann--Whitney U test statistics demonstrate that even for classical
statistical tests, the development of efficient algorithms remains an active area of research.
This highlights the need for more flexible yet efficient approaches that can be 
readily applied to a diverse range of test statistics, including custom ones
better suited to specific applied questions.

\subsection{\emph{p}-value estimation with Monte Carlo}

A general approach for calculating \emph{p}-values in permutation tests is based on Monte Carlo simulation.
When applied to our two-sample case, the idea is to randomly generate $K$ combinations $A^i \in \mathcal{P}_{=n}([n+m])$ and use them to estimate the true \emph{p}-value~(\ref{eq_prob_comb_short}) as:
\begin{equation}
    \Pr(S \ge \gamma) \approx \hat{p}_{\text{MC}} = \dfrac{1}{K} \sum_{i=1}^K [S(A^i) \ge \gamma].
\end{equation}

The drawback of the Monte Carlo approach is its high relative error~\citep{botev_efficient_2008}:
\begin{equation}
    \frac{\sqrt{\Var(\hat{p}_{\text{MC}})}}{\mathbb{E}[\hat{p}_{\text{MC}}]} = \sqrt{ \frac{1 - \Pr(S \ge \gamma \mid H_0)}{K \cdot \Pr(S \ge \gamma \mid H_0)} }.
\end{equation}
For example, to achieve a relative error of \( 100\% \) for a \emph{p}-value of \( 10^{-10} \), one would need to set \( K \approx 10^{10} \), which is computationally prohibitive.

\subsection{\emph{p}-value estimation with multilevel splitting Markov chain Monte Carlo}
\label{sec_multilevel_intro}

The multilevel splitting technique can be used to overcome the poor accuracy scaling of the classical Monte Carlo approach~\citep{glasserman_multilevel_1999}.
The general idea of splitting, when estimating the probability of an event \( \mathcal{A} \), is to consider a family of nested events \( \Omega = \mathcal{A}_0 \supset \mathcal{A}_1 \supset \ldots \supset \mathcal{A}_t = \mathcal{A} \), and compute the probability as \( \Pr(\mathcal{A}) = \prod_{i=0}^{t-1} \Pr(\mathcal{A}_{i+1} \mid \mathcal{A}_i) \).

When applied to estimating the probability \( \Pr(S \ge \gamma) \), 
we can choose a sequence of levels \( -\infty = L_0 < L_1 < \ldots < L_t = \gamma \), and consider the events \( S \ge L_i \). This leads to:
\begin{equation}
    \Pr(S \ge \gamma) = \prod_{i=0}^{t-1} \Pr(S \ge L_{i+1} \mid S \ge L_i).
\end{equation}

The task of estimating the probability \( \Pr(S \ge \gamma) \) can then be solved by
multiplying the estimates of the conditional probabilities \( \Pr(S \ge L_{i+1} \mid S \ge L_i) \).
If the levels \( L_i \) are chosen so that these conditional probabilities are relatively large,
then each of them can be accurately estimated using Monte Carlo sampling
from the corresponding conditional distribution \( \Pr(\cdot \mid S \ge L_i) \).

Sampling from the conditional distribution can be achieved using Markov chain Monte Carlo (MCMC) approaches, such as the Metropolis~\citep{metropolis_equation_1953} or Metropolis–Hastings~\citep{hastings_monte_1970} algorithms. 
For the Metropolis algorithm, we can construct a Markov chain over \( n \)-element combinations from \( [n+m] \)~\citep{korotkevich_fast_2021}.
An iteration of the Metropolis algorithm may proceed by removing a random element from the current combination and then adding another element back (possibly the same one that was removed). 
The transition is accepted if the value of the statistic \( S \) for the proposed combination satisfies the current level boundary and is rejected otherwise.
As long as all combinations satisfying the level boundary are connected, the conditions for convergence of the Metropolis algorithm are satisfied. 
For the Kolmogorov--Smirnov statistic, this connectivity can be established from the fact that the \( D^+ \) statistic can be increased for any combination, except for the one with the maximum score \( D^+ = 1 \). A similar reasoning applies to the Mann--Whitney U test.

\begin{figure*}[!t]
\centering
\includegraphics[width=\textwidth]{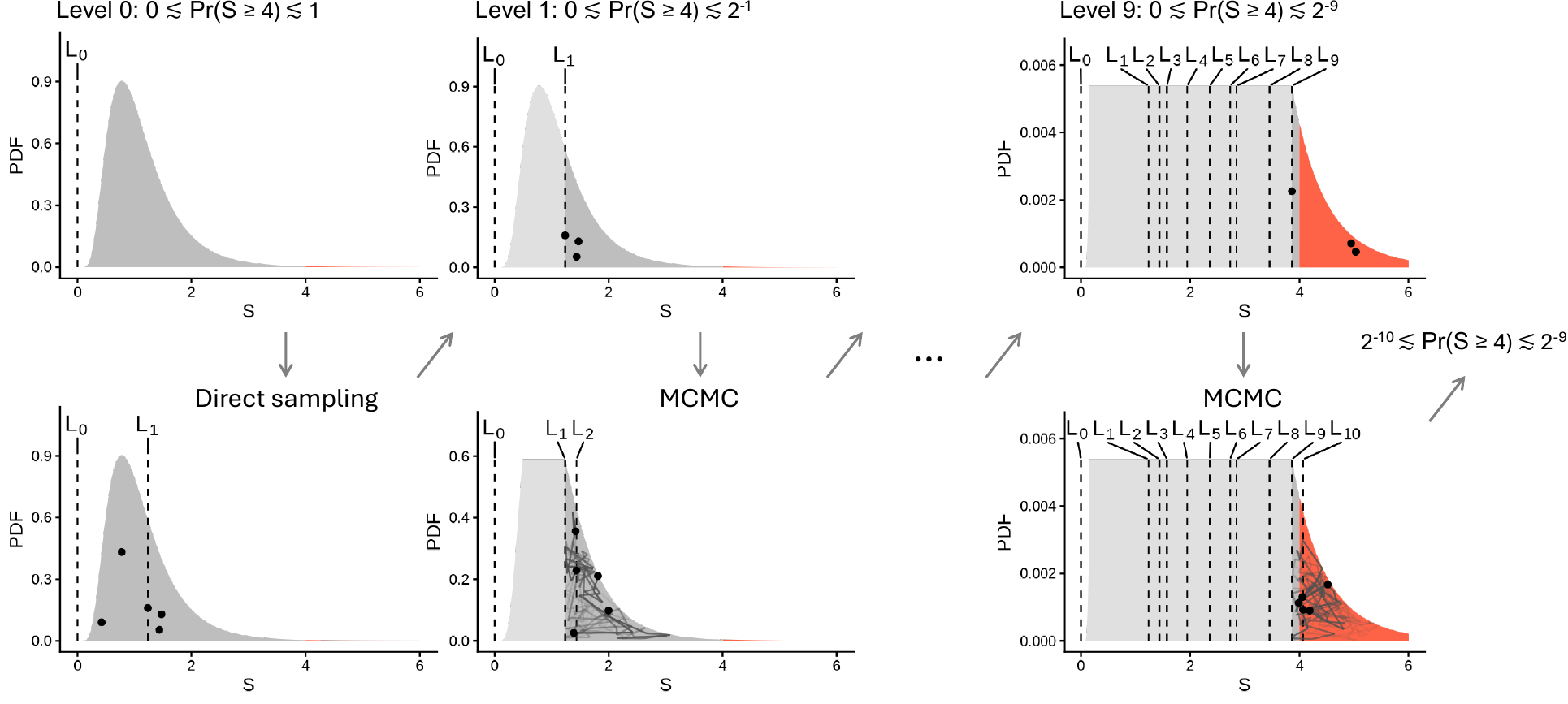}
\caption{
A toy example illustrating the use of an adaptive multilevel splitting Monte Carlo algorithm to estimate the probability \( \Pr(S \ge \gamma) \) for \( \gamma = 4 \).
Individual plots show the probability density function (PDF) of \( S \).
On each iteration, \( K = 5 \) random samples are generated, and their median score is used to define the next level, so that \( \Pr(S \ge L_i) \approx 2^{-i} \).
Iterations stop when \( L_9 \le 4 \le L_{10} \), resulting in the estimate 
\( 2^{-10} \lesssim \Pr(S \ge 4) \lesssim 2^{-9} \).}
\label{fig_multilevel_scheme}
\end{figure*}

Adaptive multilevel splitting addresses the question of how to choose the level boundaries \( L_i \)~\citep{cerou_adaptive_2007,cerou_adaptive_2019,korotkevich_fast_2021}.
Since the distribution of the statistic \( S \) is not known, selecting appropriate values in advance is not possible. 
The adaptive approach begins with a trivial level \( L_0 = -\infty \). 
A sample of size \( K \) from \( \Pr(\cdot \mid S \ge L_0) \) can be generated directly by sampling combinations from \( \mathcal{P}_{=n}([n+m]) \).
Once this initial sample is obtained, the \( \beta \)-quantile of the sampled values (e.g., \( \beta = 1/2 \)) is used to define the next level \( L_1 \). 
The samples that already exceed the new threshold \( L_1 \) are then used as starting points for an MCMC algorithm to generate samples from \( \Pr(\cdot \mid S \ge L_1) \). 
The \( \beta \)-quantile of this new sample defines the next level \( L_2 \), and the process continues iteratively.
The procedure terminates when the next level would exceed \( \gamma \). 
This adaptive scheme keeps the conditional probabilities approximately equal to \( 1 - \beta \), resulting in the estimate 
\[
(1 - \beta)^t \lesssim \Pr(S \ge \gamma) \lesssim (1 - \beta)^{t-1}.
\]
A toy example illustrating the approach is shown in Figure~\ref{fig_multilevel_scheme}.

\subsection{Multilevel splitting in the discrete case}
\label{sec_discrete_problems}

Multilevel splitting typically targets conditional probabilities to be \( 1/2 \) or greater.
Unfortunately, in the discrete setting of permutation tests,
it is not always possible to select levels that satisfy this criterion.

To formalize the above problem, first notice that the following inequality holds:
\(
\Pr(S \ge L_{i+1} \mid S \ge L_i) \le \Pr(S > x \mid S \ge x)
\)
for all \( L_i \le x < L_{i+1} \).
This implies that at least one of the conditional probabilities \( \Pr(S \ge L_{i+1} \mid S \ge L_i) \)
must be at most \( \inf_{x < \gamma} \Pr(S > x \mid S \ge x) \).

We can then define
\(
\hbar(S) = \inf_{x < \max S} \Pr(S > x \mid S \ge x)
\)
as a characteristic of the statistic \( S \), representing the most extreme jump in its distribution.
For a statistic \( S \) with \( \hbar(S) < 1/2 \), the adaptive multilevel splitting approach described in Section~\ref{sec_multilevel_intro} with \( \beta = 1/2 \) will not be able to choose an appropriate next level boundary.

We can show that for the Kolmogorov--Smirnov test with parameters \( n \le m \), the value of 
\( \hbar(D^+) \) can be as low as \( \frac{1}{n+1} \).
Consider $X = [1, \ldots, n]$, $Y = [n + 1, \ldots, n + m]$. First, note that the maximum possible value of the statistic is equal to~\( 1 \), and is achieved on the set \( A = \{1, 2, \ldots, n\} \). 
Next, from equations~(\ref{eq_def_dpos}) and~(\ref{eq_def_dpos_fg}), it follows that
the second largest value of the statistic is at most \( 1 - \frac{1}{m} \). 
This value is attained by any set of the form \( \{1, 2, \ldots, n\} \setminus \{i\} \cup \{n+1\} \). 
There are \( n \) such sets, which means that 
\[
\begin{aligned}
\hbar(D^+)
&\le \Pr(D^+ > x^* \mid D^+ \ge x^*) \\
&\le \frac{1}{n+1},
\qquad x^* = 1 - \frac{1}{m}.
\end{aligned}
\]

For the Kolmogorov--Smirnov test, large jumps in the distribution are not restricted to values of the statistic just below 1. 
They can also occur at intermediate values, where the same statistic value may be realized by many different combinations. 
This behavior becomes particularly problematic when \( \operatorname{gcd}(n, m) \) is large:
 all possible statistic values have the form \( \frac{i \operatorname{gcd}(n, m)}{nm}  \)
for \( i \in \{0, 1, \ldots, \operatorname{lcm}(n, m) \} \), and a larger \( \operatorname{gcd}(n, m) \) makes the distribution concentrate at fewer points.

Similarly, for the Mann--Whitney U test with ties we can consider the following example: sample $X = [1, 1, \ldots, 1]$, and $Y = [0, 0, \ldots, 0]$. In such a case, all the ranks in $Y$ are equal to $\frac{m + 1}{2}$, and all the ranks in $X$ are equal to $\frac{m + 1 + n + m}{2} = m + \frac{n + 1}{2}$.
In such example, the statistic $U_1 = nm + n\cdot\frac{n+1}{2}-\frac{n(n+1)}{2} = nm$, and it can be achieved by picking the highest ranks into the subset. This value is the maximum possible value of the statistic for these ranks.
The next largest value is achieved when we swap one of the zeroes with one of the ones, and is therefore equal to $nm - (m+\frac{n+1}{2})+(\frac{m+1}{2}) = nm - \frac{n+m}{2}$. The number of ways to achieve that is equal to $nm$. Therefore, 
\[
\begin{aligned}
\hbar(U_1)
&\le \Pr(U_1 > x^* \mid U_1 \ge x^*) \\
&= \frac{1}{nm + 1},
\qquad
x^* = nm - \frac{n+m}{2}.
\end{aligned}
\]

In addition to the presence of jumps in the statistic distribution, it is important to keep in mind
the general property that for discrete distributions \( \Pr(S > x) \) and \( \Pr(S \ge x) \) are distinct quantities.
Both of these issues must be taken into account when 
designing and implementing an accurate multilevel splitting scheme
for permutation-based \emph{p}-value estimation.

\section{Methods}
\label{sec_methods}

\subsection{Using hashing to improve granularity of the statistic-based combination ordering} 
\label{sec_hashing}

As discussed in Section~\ref{sec_discrete_problems}, the straightforward application 
of the multilevel splitting scheme to permutation tests is complicated by
cases where many combinations have the same value of the statistic.
To address this issue, we propose introducing an additional criterion for ordering the 
combinations.

More formally, let us consider a strict total order \( A \succ_S B \) such that
\(
S(A) > S(B) \implies A \succ_S B
\).
In this case, we can use specific combinations as level boundaries instead of 
statistic values and rewrite the target probability as:
\begin{equation}
\label{eq_prob_comb_geq}
\begin{aligned}
\Pr(S(X) \ge \gamma)
= & \Pr(S(X) \ge \gamma \mid X \succeq_S L_t) \\
&\cdot \prod_{i=0}^{t-1}
\Pr(X \succeq_S L_{i+1} \mid X \succeq_S L_i) ,
\end{aligned}
\end{equation}
where \( X \in \mathcal{P}_{=n}([n+m]) \) is a random \( n \)-combination, \( L_i \in \mathcal{P}_{=n}([n+m]) \) are the level boundary combinations for $i > 0$,
and $L_0$ is a special object such that $X \succ_S L_0$ is always true---an equivalent of $-\infty$.

With such a total order in hand, we can guarantee that for any level \( L_i \), there exists 
a combination \( L_{i+1} \succ_S L_i \) such that
\(\Pr(X \succeq_S L_{i+1} \mid X \succeq_S L_i) \geq 1/2\) by choosing the next combination after $L_{i}$ with respect to $\succ_S$.
The only exception is one of the combinations where the maximal possible value of the statistic \( S \) is attained.

Note that the required order can be constructed by introducing a secondary criterion for comparing combinations.
A simple example is to use lexicographic comparison, defined as:
\begin{equation}
\begin{aligned}
A \succ_{S,\mathrm{lex}} B
:=\;& \bigl(S(A) > S(B)\bigr) \\
&{}\OR
\bigl(S(A) = S(B) \AND A >_{\mathrm{lex}} B\bigr),
\end{aligned}
\end{equation}
where \( >_{\mathrm{lex}} \) denotes lexicographic ordering of combinations.

A more practical approach, however, is to use an \( s \)-bit hash function \( h \) that maps
each combination in \( \mathcal{P}_{=n}([n+m]) \) to an integer in the range \( [0, 2^s) \). 
Using this function, we can define the \( \succ_{S,h} \) operator as:
\begin{equation}
\begin{aligned}
A \succ_{S,h} B
:=\;& \bigl(S(A) > S(B)\bigr) \\
&{}\OR
\bigl(S(A) = S(B) \AND h(A) > h(B)\bigr).
\end{aligned}
\end{equation}
While \( \succ_{S,h} \) is not necessarily a true total order due to potential
collisions, an \( s \)-bit hash function can resolve conditional probability jumps as small as \( 2^{-s} \).

The practical advantage of using a hash function over lexicographic comparison is that 
a hash function can be designed to allow \( O(1) \) updates during MCMC steps and \( O(1) \) comparisons. 
For example, let \( H_i \), \( i \in [n+m] \), be independent samples from the discrete uniform distribution on \( [0, 2^s) \). 
Then the hash function can be defined as:
\begin{equation}
    h(A) = \bigoplus_{i \in A} H_i.
\end{equation}
This function can be efficiently updated when a single element is 
replaced in a combination, since \( h(A') = h(A) \oplus h(A \oplus A') \),
where \( A \oplus A' \) denotes the symmetric difference between the combinations \( A \) and \( A' \).

\subsection{Proposed adaptive multilevel splitting Monte Carlo algorithm}

Leveraging the \( \succ_{S,h} \) ordering defined in Section~\ref{sec_hashing},
we propose the following hash-augmented adaptive multilevel splitting Monte Carlo algorithm (Algorithm~\ref{alg_multilevel}) for estimating two-sample permutation test \emph{p}-values.
The algorithm follows the general adaptive multilevel splitting scheme described in Section~\ref{sec_multilevel_intro}.
Each iteration begins with a sample from the conditional distribution
at the current level. 
This sample is used to select a boundary for the next level.
The probability estimate is then updated to reflect the introduction of the new level.
Finally, an MCMC-based procedure is used to generate a sample from the conditional distribution at the new level.
The iterations stop when the observed value of the statistic is reached or exceeded.
However, this algorithm introduces several distinctive features and potential modifications,
which we discuss next.

\begin{algorithm}[!htb]
\caption{Hash-augmented adaptive multilevel splitting Monte Carlo for estimating two-sample permutation test \emph{p}-values}
\label{alg_multilevel}
\begin{algorithmic}[1]
\Statex \textbf{Input:} \(n\), \(m\) -- sizes of the samples
\Statex \hspace{\algorithmicindent}\(K\) -- number of Monte Carlo samples
\Statex \hspace{\algorithmicindent}\(S\) -- a statistic function that operates on subsets of size \(n\) from \([n+m]\)
\Statex \hspace{\algorithmicindent}\(h\) -- a hash function that operates on the same domain as \(S\)
\Statex \hspace{\algorithmicindent}\(\gamma\) -- the value of the statistic \(S\) for which \emph{p}-value should be calculated
\Statex \textbf{Result:} \(\log \hat{p}\) -- estimate for \(\log \Pr(S \ge \gamma \given H_0)\)
\Statex \hspace{\algorithmicindent}\(\hat{\sigma}^2\) -- variance estimate
\For{\(i \leftarrow 1\) \textbf{to} \(K\)}
    \State \(A^i \gets\) generate a random \(n\)-combination from \([n+m]\)
\EndFor
\State \(\log \hat{p} \gets 0\); \(\hat{\sigma}^2 \gets 0\)
\For{\(j \leftarrow 0\) \textbf{to} \(\infty\)}
    \State \(\triangleright\) Invariant: \(A\) is a sample from \(\Pr(X \given X \succ_{S,h} L_j)\)
    \State \(L_{j+1} \gets\) select the next boundary element from \(\{A^i\}_{i=1}^K\) \label{algln_select_boundary}
    \If{\(S(L_{j+1}) \ge \gamma\)}
        \State \(M_j \gets \#\{i \mid S(A^i) \ge \gamma\}\)
    \Else
        \State \(M_j \gets \#\{i \mid A^i \succ_{S,h} L_{j+1} \} + 1\) \label{algln_ml_ord}
    \EndIf
    \State \(\log \hat{p} \gets \log \hat{p} + \psi(M_j) - \psi(K+1)\) \label{algln_ml_logp}
    \State \(\hat{\sigma}^2 \gets \hat{\sigma}^2 + \psi_1(M_j) - \psi_1(K+1)\) \label{algln_ml_sigma}
    \If{\(S(L_{j+1}) \ge \gamma\)}
        \State \textbf{break}
    \EndIf
    \State \(A \gets \textsc{Perturb}(A, \cdot \succ_{S,h} L_{j+1})\)
    \Statex \hspace{\algorithmicindent}\(\triangleright\) makes \(\forall i: A^i \succ_{S,h} L_{j+1} \)
\EndFor
\end{algorithmic}
\end{algorithm}

First, the level boundaries are defined not by the statistic values, but by specific \( n \)-combinations. 
Further, the algorithm uses conditional probabilities of the form \( \Pr(X \mid X \succ_{S,h} L) \)
instead of \( \Pr(X \mid X \succeq_{S,h} L) \) as used in~(\ref{eq_prob_comb_geq}).
While both formulations are equivalent in the continuous multilevel splitting setting,
the former is preferred in discrete settings, as discussed in Section~\ref{sec_perturb}.

Consequently, the selection of the next level boundary (line~\ref{algln_select_boundary})
is performed at the level of sampled combinations. The boundary can be chosen as the median
or any other quantile (with respect to the \( \succ_{S,h} \) order) of the current sample.
However, it is important to additionally ensure that the new boundary differs from the current level boundary,
and that the next level is non-empty---that is, at least one element in the current sample satisfies the new level boundary.
These properties can be violated in a direct implementation due to the presence of duplicate combinations in the current level sample.
The final level contains only a single element and   must be handled explicitly.

Since the boundary does not always correspond to a predefined order statistic,
the conditional probabilities for each level must be estimated directly from the observed samples. 
To this end, we compute a value \( M_j \) (line~\ref{algln_ml_ord}) such that 
\(\Pr(X \succ_{S,h} L_{j+1} \mid X \succ_{S,h} L_j) \approx M_j / (K+1)\).
If \( \succeq_{S,h} \) is used instead to define the level condition, then line~\ref{algln_ml_ord}
should be replaced with \( M_j \gets \# \{i \mid A_i \succeq_{S,h} L_{j+1}\} \).

Following the approach we previously proposed in~\cite{korotkevich_fast_2021}, 
the algorithm estimates the logarithm of the desired \emph{p}-value.
This is reflected in lines~\ref{algln_ml_logp} and~\ref{algln_ml_sigma},
where \( \psi \) denotes the digamma function and \( \psi_1 \) the trigamma function.
The final estimate can be interpreted as the sum of logarithms of \( M_j \)-order statistics from a \( K \)-sample drawn from the standard uniform distribution \( U(0, 1) \).
Order statistics from the uniform distribution follow a Beta distribution,
and the expectation and variance of their logarithms can be expressed in terms of \( \psi \) and \( \psi_1 \), respectively.
The resulting estimate is approximately normally distributed, centered at the true value, with variance \( \sigma^2 \).
Consequently, the interval \( (\log \hat{p} - 2\hat{\sigma},\, \log \hat{p} + 2\hat{\sigma}) \) 
can be used as an approximation of a 95\% confidence interval.

Finally, the \texttt{Perturb} function implements a Markov chain Monte Carlo-based algorithm
to generate a sample for the next level. We discuss it in more detail in the next section.

\subsection{Sampling from the conditional distributions}
\label{sec_perturb}

Algorithm~\ref{alg_multilevel} critically depends on the ability to obtain 
independent samples
from the conditional distributions \( \Pr(X \mid X \succ_S L_j) \).
This task can be challenging, but it can be approached using a version of the Markov chain Monte Carlo (MCMC) algorithm, 
such as the Metropolis algorithm mentioned in Section~\ref{sec_multilevel_intro}.
Theoretically, for an appropriate Markov chain, the Metropolis algorithm converges 
to the desired distribution, given enough iterations regardless of the initial state.
However, in practice, we aim to minimize the number of Metropolis iterations 
in order to improve the overall runtime of the algorithm.
This is the goal of Algorithm~\ref{alg_perturb}.

\begin{algorithm}[!htb]
\caption{Generating a sample from $\Pr(X \given X \succ_S L_{j+1})$}
\label{alg_perturb}
\begin{algorithmic}[1]
\Statex \textbf{Input:} \(n\), \(m\) -- sizes of the samples
\Statex \hspace{\algorithmicindent}\(K\) -- number of Monte Carlo samples
\Statex \hspace{\algorithmicindent}\(A^i\) -- samples from \(\Pr(X \given X \succ_S L_{j})\)
\Statex \hspace{\algorithmicindent}\(L_{j+1}\) -- next level boundary
\Statex \hspace{\algorithmicindent}\(\mathrm{type}\) -- resampling type: \(\mathrm{full}\) (default) or \(\mathrm{partial}\)
\Statex \hspace{\algorithmicindent}\(\alpha\) -- scaling parameter (default: 1)
\Statex \textbf{Result:} \(B^i\) -- samples from \(\Pr(X \given X \succ_S L_{j+1})\)
\State \(A' \gets (A^i \mid A^i \succ_S L_{j+1})\)
\State \(B \gets A'\)
\While{\(|B| < K\)}
    \State \(B \gets (B, \text{random element of } A')\)
\EndWhile
\State \(r \gets \begin{cases}
    1 & \text{if }\mathrm{type} = \mathrm{full}\\
    |A'|+1 & \text{if } \mathrm{type} = \mathrm{partial}
\end{cases}\)
\State \(n_{\text{ac}} \gets 0\), \(n_{\text{it}} \gets 0\)
\While{\( n_{\text{\upshape ac}} / (K-r+1) < \alpha n / 2 \)}
    \For{\(i \leftarrow r\) \textbf{to} \(K\)}
        \State \(T \gets B^i \setminus \{ \text{random element of } B^i \} \cup \{ \text{random element of } [n+m] \}\)
        \If{\( |T| = n \AND T \succ_S L_{j+1} \)}
            \State \(B^i \gets T\)
            \State \(n_{\text{ac}} \gets n_{\text{ac}} + 1\)
        \EndIf
    \EndFor
    \State \(n_{\text{it}} \gets n_{\text{it}} + 1\)
\EndWhile
\For{\(j \leftarrow 1\) \textbf{to} \(n_{\text{\upshape it}}\)}
    \For{\(i \leftarrow r\) \textbf{to} \(K\)}
        \State \(T \gets B^i \setminus \{ \text{random element of } B^i \} \cup \{ \text{random element of } [n+m] \}\)
        \If{\( |T| = n \AND T \succ_S L_{j+1} \)}
            \State \(B^i \gets T\)
        \EndIf
    \EndFor
\EndFor
\end{algorithmic}
\end{algorithm}

First, note that before generating \( K \) samples from \( \Pr(X \mid X \succ_S L_{j+1}) \),
we already have \( K \) samples \( A^i \) from \( \Pr(X \mid X \succ_S L_j) \),
obtained either in the previous iteration or by direct sampling before the first iteration. 
Some of these samples also satisfy the condition for the next level:
\( A' := (A^i \mid A^i \succ_S L_{j+1}) \), and \(A'\) approximately constitutes 
an independent identically distributed (i.i.d.) sample
from the next-level conditional distribution \(  \Pr(X \mid X \succ_S L_{j+1}) \).
Importantly, unlike in the continuous case, 
even if we assume that \( ( A^i )  \) is an i.i.d. sample
from \( \Pr(X \mid X \succ_S L_{j}) \) the subset \( A' \)
is not a true i.i.d. sample from \( \Pr(X \mid X \succ_S L_{j+1}) \)
because \( L_{j+1} \) was selected from \( A^i \) as an order statistic and
discrete order statistics lack the Markov property~\citep{nagaraja_non-markovian_1982,dembinska_discrete_2014}.
This issue, however, is smoothed by using a total (or close to total) 
order when the support of \( \Pr(X \mid X \succ_S L_{j}) \) is large
and probability of ties is low.
Notably, the \( \succeq_S \) variant introduces an additional bias
due to the deterministic inclusion of the boundary value \( L_{j+1} \) into the sample.
Nevertheless, in both \( \succ_S \) and \( \succeq_S \) variants
these samples can be used as starting points for the Metropolis algorithm.

Given that we have \( M_j - 1 > 0 \) samples from \( \Pr(X \mid X \succ_S L_{j+1}) \),
there are two main options. 
The first is to use the Metropolis algorithm to generate \( K - M_j + 1 \) new independent samples. 
The second is to randomly duplicate the existing samples until their count reaches \( K \), 
and then apply Metropolis iterations to all of them.
We refer to the first approach as \emph{partial resampling}, 
and the second as \emph{full resampling}.

Partial resampling is a natural and commonly used strategy in adaptive multilevel splitting~\citep{cerou_adaptive_2019}. 
An additional advantage is that its computational efficiency can be tuned 
by selecting different target values for the conditional probability 
\( \Pr(X \succ_S L_{j+1} \mid X \succ_S L_j) \).
Higher values of this probability (up to \( K / (K + 1) \)) increase 
the total number of levels, but improve the ratio of estimation precision 
to the number of newly generated samples.

Full resampling is less standard, but it is more robust to various sampling biases, 
since all samples undergo the same number of Metropolis iterations over the same Markov chains. 
For example, it is compatible with sampling from the \( \Pr(X \mid X \succeq_S L_{j+1}) \) distribution.
This type of resampling was previously proposed and tested by our group in the 
FGSEA-multilevel method~\citep{korotkevich_fast_2021}.
Since full resampling always applies Metropolis steps to all \( K \) samples, 
unlike in partial resampling, there is no benefit in targeting higher values of the conditional probabilities. 
Consequently, this resampling strategy is best used with level selection based on the median of the samples.

Regardless of the resampling type, a major practical question is when to stop the Metropolis iterations. 
A particularly challenging aspect is that different levels operate over different Markov chains, 
which may have varying mixing times. The mixing time also depends on the specific statistic \( S \) being used.

Nevertheless, we propose the following scheme, parameterized by \( \alpha \), 
as an improvement over the approach used in the FGSEA-multilevel algorithm~\citep{korotkevich_fast_2021}. 
The idea is to aim for \( \alpha n \) accepted transitions in the Metropolis steps for each sample, 
ensuring that the samples move sufficiently along the Markov chain regardless of the level.
However, stopping the Metropolis steps individually for each sample after \( \alpha n \) accepted 
transitions would introduce sampling biases. Instead, we adopt a two-stage resampling procedure. 
In the first stage, Metropolis steps are run in parallel for all samples until the average number of 
accepted transitions reaches \( \alpha n / 2 \). 
In the second stage, the same number of additional Metropolis steps is performed for each sample, 
regardless of whether the transitions are accepted or rejected.

\subsection{Implementation considerations}

There are several important practical details that need to be considered when implementing
the above algorithm.

First of all, an important aspect of the implementation is the precision of the statistic calculation. 
The proposed algorithm relies on the ability to exactly compare the values of the statistic 
\( S(A) \) and \( S(B) \) for two combinations \( A \) and \( B \).
However, if floating-point arithmetic is used, the computed values of the statistic may differ slightly 
due to precision errors, making comparisons between \( S(A) \) and \( S(B) \)
unstable for close or equal values. 
To address this, we assume the statistic has integer values, and implement the statistic calculation and comparison 
using integer arithmetic. For example, for the Kolmogorov--Smirnov test we multiply all values by the common denominator $nm$.

In addition, performance benefits from efficient updates of 
the statistic value during Metropolis steps, as it is the hot spot of the algorithm. For example, for the Mann--Whitney U test, one can store the current value of the statistic and recalculate it by adding the difference of ranks of the elements being swapped, resulting in constant time recalculation, compared to linear time if the score is calculated each time from scratch.

Lastly, two-sided test $p$-values can commonly be computed as a sum of two one-sided $p$-values, thus being compatible with the proposed algorithm. However, for the Kolmogorov--Smirnov test, the two-sided test is defined as
\( \Pr(D \ge \gamma \mid H_0) \)
for a distinct statistic \(
D = \sup_x \left | F(x) - G(x) \right |
\).
This statistic is not directly compatible with the algorithm as the Markov chain becomes disconnected: then the maximum value $D=1$ is attained by both combinations $A_1=\{1, 2, \ldots, n\}$ and $A_2=\{m+1,m+2,\ldots,m+n\}$, which cannot be transformed into each other without decreasing the value.





\subsection{\texttt{hamstest} package}
\label{sec_software}

We developed \texttt{hamstest} (\url{https://pypi.org/project/hamstest/}), a Python package that implements the proposed
algorithm to calculate permutation \emph{p}-values for two-sample statistical
tests with user-defined test statistics.
In the package, the test statistic is defined by a computational procedure taking
an $(n, m)$-partition as an input, and is used to
define the~\(\succ_{S}\) relationship for subset comparisons
in the multilevel splitting Monte Carlo algorithm.
The computation procedure alongside the algorithm parameters
(scaling parameter $\alpha$, number of samples $K$, resampling variant)
is used to construct an \texttt{Estimator} object,
whose \texttt{estimate} method can then be used to calculate a \emph{p}-value
for a given statistic value.

For higher efficiency, the package provides two major optimization options
to reduce the time taken by MCMC iterations.
The first option is to provide an additional function to recalculate
the score when two elements are swapped in the $(n, m)$ partition. 
For example, for the Mann--Whitney U statistic such recalculation
can be done in $O(1)$ time.
The second option is to implement the scoring calculation and recalculation procedures
in C++, in which case the main MCMC loop is executed by compiled code.

As an example of using the package, we provide a reference implementation of the Mann--Whitney U test.
It demonstrates the use of both Python and C++ interfaces, as well as specifying
the score recalculation procedure.

\section{Results}

\subsection{Evaluation of returned confidence intervals}
\label{sec_results_ci}

First, we evaluated the practical convergence of the proposed hash-augmented adaptive multilevel splitting Monte Carlo algorithm 
for \emph{p}-value estimation in Kolmogorov--Smirnov and Mann--Whitney U tests by assessing the quality of the returned 
confidence intervals. 
To do this, we considered the task of estimating the \emph{p}-value 
\( \Pr(S \geq S_\text{max}) \), 
where \( S_\text{max} \) is the maximal possible value of the statistic: 1 in 
the case of the Kolmogorov--Smirnov test and \(n m \) for the Mann--Whitney U test.
For this case, the true \emph{p}-value has the closed form \( 1 / \binom{n+m}{n} \). 
We considered a fixed value \( n + m = 1000 \) and values of \( n \in \{50, 100, 250\} \),
in which case the target \emph{p}-values were equal to 
\( 1.1 \cdot 10^{-85} \), \( 1.6 \cdot 10^{-140} \), and 
\( 2.1 \cdot 10^{-243} \) for \( n = 50 \), 100, and 250, respectively.
We expect the interval 
\( (\log \hat{p} - 2\hat{\sigma}, \log \hat{p} + 2\hat{\sigma}) \) 
to approximate a 95\% confidence interval. 
Thus, we evaluate the fraction of independent runs in which the true value 
\( \log \Pr(S \geq S_\text{max}) \) falls within this interval.

Three versions of the algorithm are considered:
\begin{enumerate}
    \item \emph{full-median} -- using full resampling when selecting the next boundary as the median of the current samples,
    \item \emph{partial-median} -- similar to the previous one but with partial resampling, 
    \item \emph{partial-min} -- also using partial resampling, but selecting the minimum of the current samples for the next boundary.
\end{enumerate}
All variants were run with varying values of the scaling parameter~$\alpha$.
The number of Monte Carlo samples generated at each level was set to $K = 101$.

\begin{figure*}[!t]
\centering
\includegraphics[width=\textwidth]{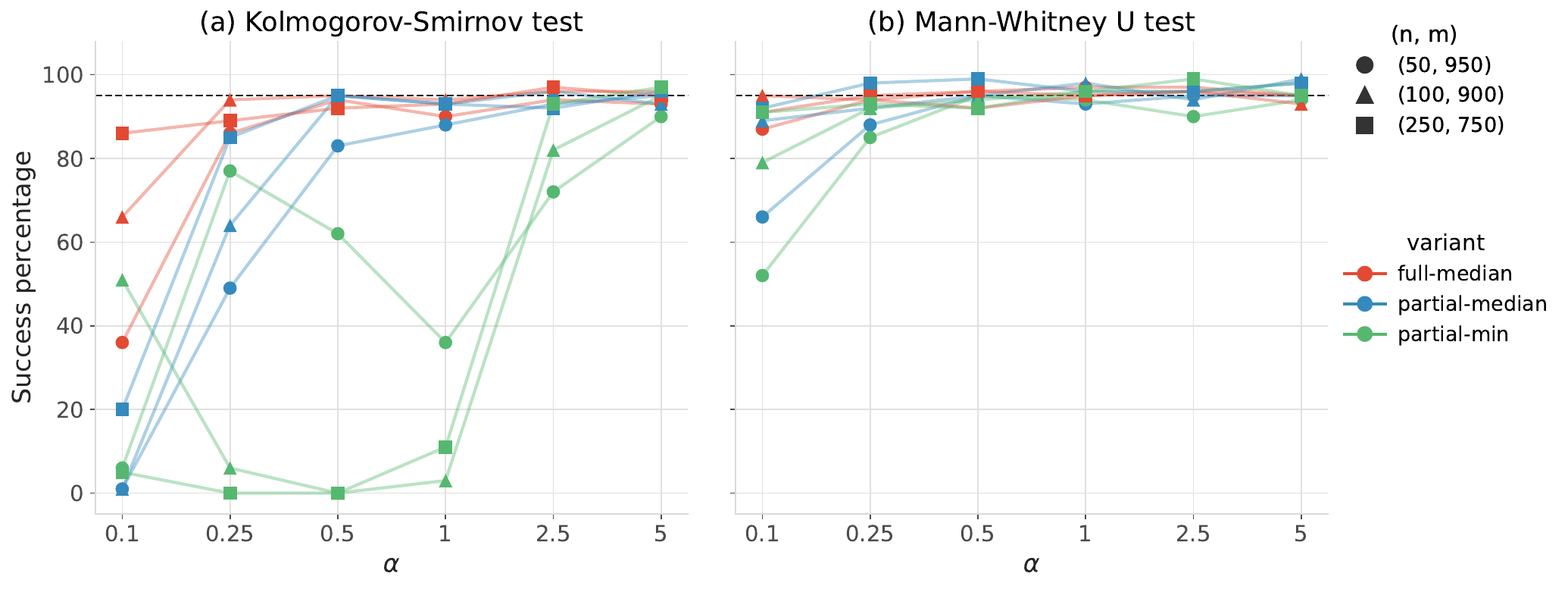}
\caption{Fraction of runs in which the 
true value of \( \log \Pr(D^+ \geq 1) \) (\emph{a})
or \( \log \Pr(U_1 \geq nm) \) (\emph{b})
falls within the estimated interval
\( (\log \hat{p} - 2\hat{\sigma},\, \log \hat{p} + 2\hat{\sigma}) \)
as a function of the scaling parameter \( \alpha \) and the type of resampling procedure. 
The three variants are: 
\emph{full-median} — full resampling with median boundary selection; 
\emph{partial-median} — partial resampling with median boundary selection; 
\emph{partial-min} — partial resampling with minimum boundary selection. 
Each point corresponds to 100 independent runs.}
\label{fig_ci_alpha_1000}
\end{figure*}

For the Kolmogorov--Smirnov test, we observe slightly different convergence behaviors
of the tested algorithm versions with the increase of \( \alpha \) (Figure~\ref{fig_ci_alpha_1000}~(a)). 
The \emph{full-median} version achieves the fastest convergence at \(\alpha \gtrsim  0.5\),
the \emph{partial-median} version starts converging from \(\alpha \gtrsim  2.5\), and 
the \emph{partial-min} version converges after \(\alpha \gtrsim  5\).
For the Mann--Whitney U test, all variants converge much faster (Figure~\ref{fig_ci_alpha_1000}~(b)): the \emph{full-median} variant starts converging after  \( \alpha \gtrsim 0.25 \)
and both \emph{partial-median} and \emph{partial-min} converge after \( \alpha \gtrsim 0.5 \).

The above results imply that the full resampling variant should be preferred in practice.
While the \emph{partial-min} version requires generating the least number of samples
from conditional distributions \(\Pr(X \mid X \succ_S L_j)\) for a fixed relative error, 
it can require significantly more internal MCMC iterations. The \emph{partial-median}
version requires half as many samples as \emph{full-median} but also requires more
MCMC iterations for convergence. 
In addition, as discussed in Section~\ref{sec_perturb}, full resampling is more robust to sampling biases. 
Therefore, we recommend using the full resampling variant as the default option, 
and suggest a default value \( \alpha = 1 \) for the scaling parameter.

\subsection{Evaluation of estimate accuracy across different input parameters}
\label{sec_results_accuracy}

Next, we evaluated the accuracy of the proposed algorithm across different values of \( n \), \( m \), 
and the true \emph{p}-value. The total sample size \( n + m \) was varied in \( \{1000, 2500, 5000, 10000\} \), 
and the value of \( n \) was varied in \( \{50, 100, 250, 500\} \).

Here, the experiments for Kolmogorov--Smirnov and Mann--Whitney U tests are different. For the Kolmogorov--Smirnov test, for each pair \( n, m \), a score value \( D^+ \) was selected
so that the true \emph{p}-value is approximately \( 10^{-10} \), \( 10^{-50} \), and \( 10^{-100} \), if such a value existed. 
True \emph{p}-values were calculated using the exact algorithm described in~\cite{korotkevich_fast_2021}, 
originally developed for the GSEA statistic but naturally supporting
the Kolmogorov--Smirnov statistic too, which provides accurate estimates even far below machine epsilon. For the Mann--Whitney U test, there is no computationally efficient exact algorithm, so instead of considering different \emph{p}-values we considered only the most extreme case $U_1 = nm$ for each pair \( n, m \), computing the corresponding exact \emph{p}-value through combination numbers as in Section~\ref{sec_results_ci}.

In this analysis, for both tests we fixed the resampling variant to full, the scaling parameter to \( \alpha = 1 \), and the number of Monte Carlo samples to $K = 101$.
For each combination of input parameters, 100 independent estimation runs were performed.

\begin{figure*}[!t]
\centering
\includegraphics[width=\textwidth]{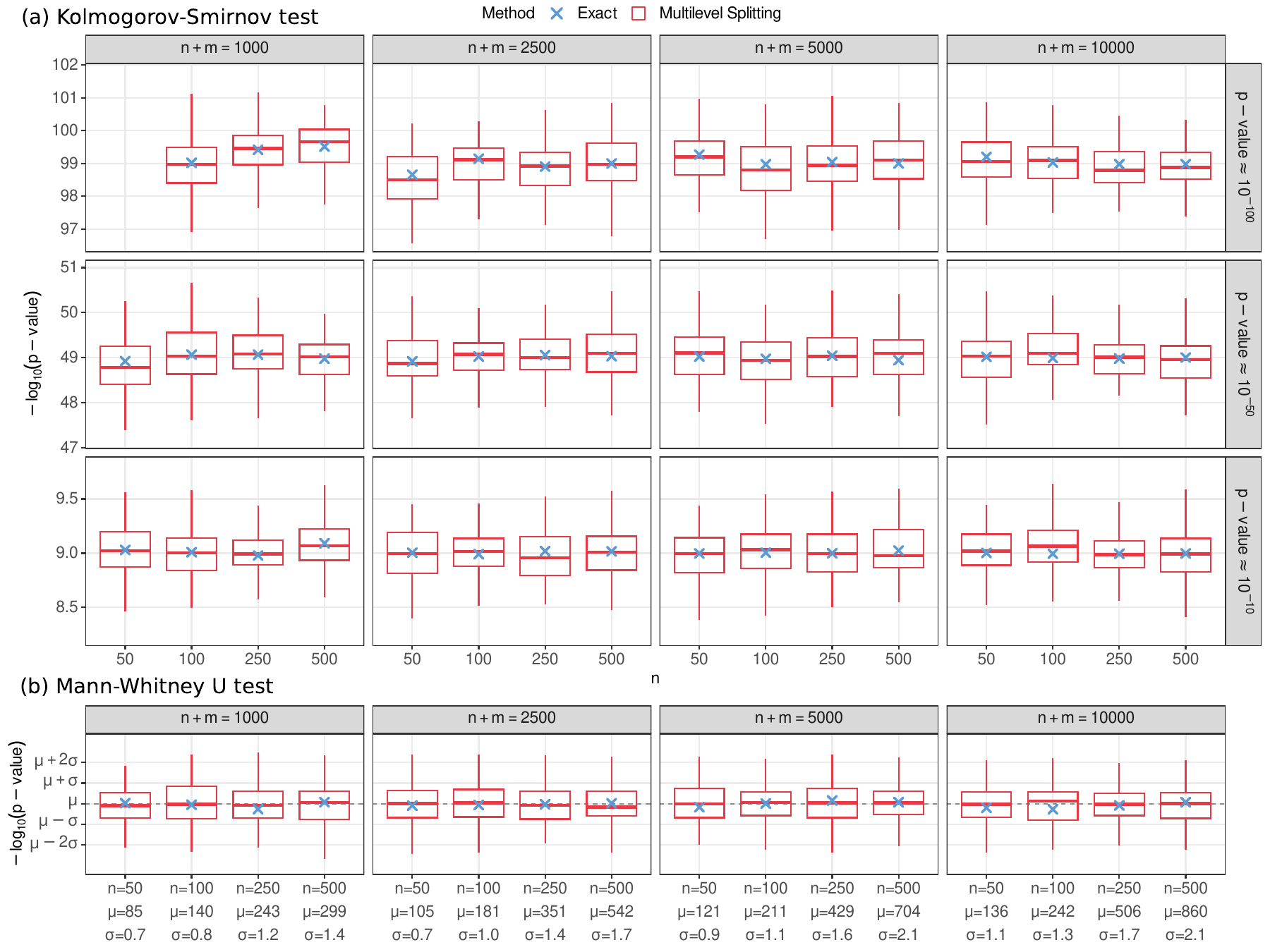}
\caption{Results of \emph{p}-value estimation computed using the proposed adaptive multilevel splitting scheme for the Kolmogorov--Smirnov test (\emph{a}) and for the Mann--Whitney U test (\emph{b}). 
Each box plot represents 100 independent runs. Crosses correspond to values calculated exactly. For the Mann--Whitney U test, \( \mu \) and \( \sigma \)
are mean and standard deviation values calculated from the estimations.}
\label{fig_accuracy}
\end{figure*}

The results of this analysis are shown in Figure~\ref{fig_accuracy}. 
For both tests, the estimates are highly consistent with the true values for all sets of parameters, 
with the median of the estimates closely matching the true \emph{p}-value. 
As expected, the spread of the estimates depends primarily on the magnitude of the true \emph{p}-value 
and not on the values of \( n \) and \( m \). 
No visible bias is observed in the estimates, supporting the conclusions 
from Section~\ref{sec_results_ci}.

\subsection{Running time of the proposed algorithm}

Finally, we analyzed the running time of the proposed algorithm. 
For both tests we ran estimation of $p$-values for the maximal
possible statistic value, but with an early exit when the current estimate
reached \( 10^{-10} \), \( 10^{-25} \), \( 10^{-50} \), or \( 10^{-100} \).
We used the same combinations of $n$ and $m$ as in the previous section. 
The algorithm settings were also identical: 
the resampling variant was set to full, the scaling parameter to \( \alpha = 1 \), 
and the number of Monte Carlo samples to \( K = 101 \). 
Timing was measured on an Apple M3 Pro processor at 4 GHz.

\begin{figure*}[!t]
\centering
\includegraphics[width=\textwidth]{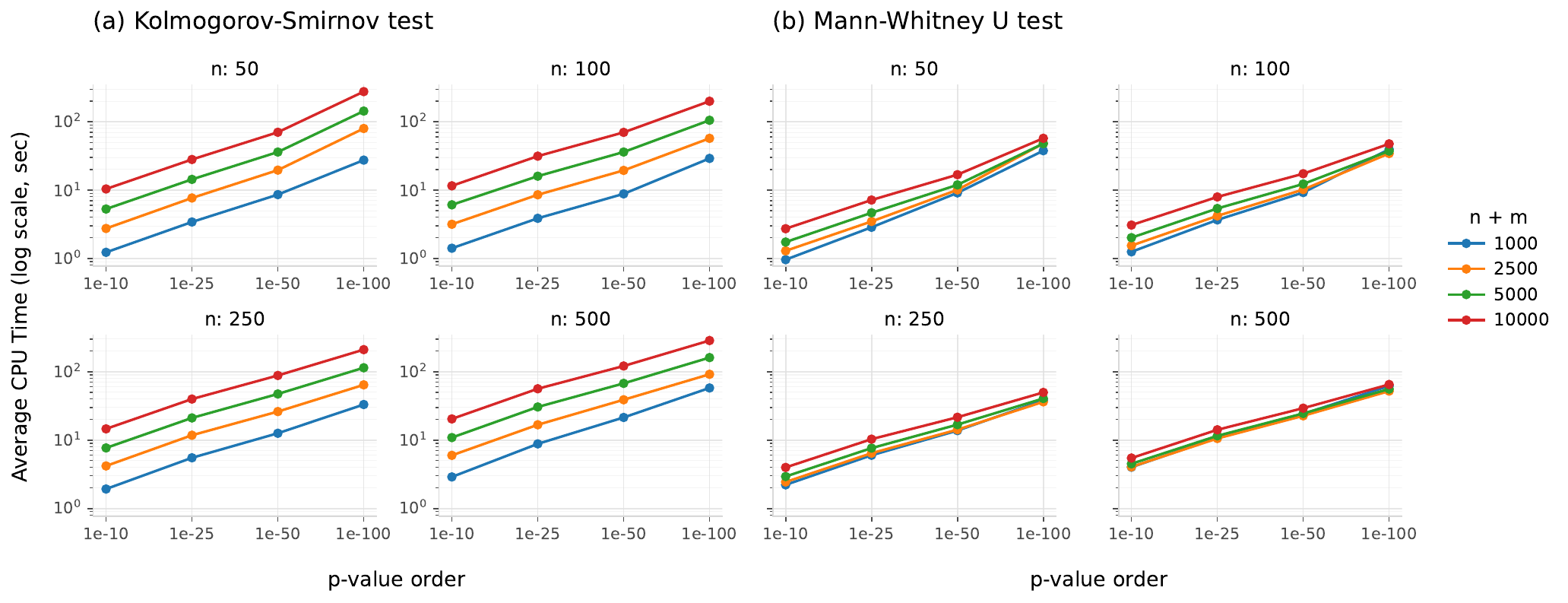}
\caption{
Running time of the proposed adaptive multilevel splitting algorithm for \emph{p}-value estimation in Kolmogorov--Smirnov (\emph{a}) and Mann--Whitney U (\emph{b}) tests. Each point represents an average over 10 runs.}
\label{fig_running_time}
\end{figure*}

Figure~\ref{fig_running_time} shows the running times. 
As expected, for the Kolmogorov--Smirnov statistic, the running time depends primarily on the order of magnitude of the \emph{p}-value and 
the value of \( n + m \). The dependence on \( n + m \) arises from the cost of recalculating 
the statistic. The dependence on the \emph{p}-value is twofold: lower \emph{p}-values 
increase the number of levels, and higher levels require more Metropolis steps 
due to lower acceptance rates. For the Mann--Whitney U test the dependence on \( n + m \) is less significant, as the cost of recalculation of the score is constant.

\section{Discussion}

In this work, we described a hash-augmented adaptive multilevel splitting Monte Carlo 
algorithm for the estimation of arbitrarily small \emph{p}-values in two-sample permutation tests. 
The algorithm is specifically designed to accurately handle challenges 
arising from the discreteness of the null distribution in such tests.

A central challenge addressed by our algorithm is the case where all samples 
at a particular level share the same statistic value, 
making it impossible to define a new level based on the statistic alone. 
The Kolmogorov--Smirnov statistic is an example where this situation is particularly
problematic: for sample sizes $n, m$ with a larger  \( \operatorname{gcd}(n, m) \),
the null distribution is concentrated at fewer points, leading to large jumps
in \emph{p}-values.
In practice, we observed this behavior in applications of 
the FGSEA-multilevel method\footnote{\url{https://github.com/alserglab/fgsea/issues/151}}.
To overcome this, we introduced the use of hash values as a secondary criterion 
for comparing combinations. This approach enables a practically total ordering even in the presence 
of repeated statistic values and can be universally applied to ensure stable level selection 
for arbitrary statistics.

Furthermore, we demonstrated high practical convergence of the developed algorithm 
to the true value while producing accurate confidence intervals. 
In our experience, with the partial resampling scheme, even small biases 
can be amplified across multiple levels, leading to systematically shifted estimates. 
In contrast, full resampling is more robust to biases in Markov chain Monte Carlo sampling, 
as it perturbs all samples continuously. 
Nevertheless, even with full resampling, a precise implementation of the algorithm 
improves the accuracy of estimates while requiring fewer 
Markov chain Monte Carlo steps.

We designed the algorithm to be generally applicable to two-sample
permutation test statistics, and the experiments confirm its convergence and accuracy 
for structurally different Kolmogorov--Smirnov and Mann--Whitney U statistics.
However, the stopping rule for the Metropolis steps may require adjustments
depending on the specific properties of the statistic.
Nevertheless, we expect that the full resampling procedure with \( \alpha = 1 \)
should be a good default choice for many types of statistics.
For any given statistic, an analysis of confidence intervals—similar to that 
presented in Section~\ref{sec_results_ci}—can be performed to assess convergence, 
as it does not rely on the availability of an exact algorithm for comparison.

Another aspect of the Metropolis stopping rule is its relationship with the mixing 
time of the corresponding Markov chain. The difference in the values of \( \alpha \) 
required for convergence, as shown in Figure~\ref{fig_ci_alpha_1000}, 
suggests that the full resampling scheme requires significantly fewer steps 
than the nominal mixing time. 
This observation is practically important, as it enables more time-efficient computation, 
and may represent an interesting direction for future research.

To conclude, the proposed adaptive multilevel sampling
algorithm provides a major improvement over 
the classical Monte Carlo approach for \emph{p}-value estimation,
while being almost as flexible when applied to different statistics.
A previous version of this algorithm has already proven successful in bioinformatics:
when applied to 
GSEA pathway enrichment analysis~\citep{korotkevich_fast_2021} and 
in the gene signature-based search engine for public gene expression 
datasets~\citep{sukhov_coresh_2025}.
The hash augmentation and other techniques described here address the problems
inherent to permutation tests because of their discreteness
and make the algorithm applicable to a broad class of two-sample statistics, 
provided that the underlying Markov chain, induced by the statistic,
is connected and mixes adequately.
The \texttt{hamstest} package provides an accurate implementation of the algorithm
and can be used to calculate \emph{p}-values for user-specified statistics.

\section{Availability}
\label{sec_software_repro}

The package \texttt{hamstest} is available at PyPI: \url{https://pypi.org/project/hamstest/}, its source code is available on GitHub \url{https://github.com/golikov-nik/hamstest}. 
Source code to reproduce results in 
Figures~\ref{fig_ci_alpha_1000}--\ref{fig_running_time}
is available at \url{https://github.com/golikov-nik/hamstest-paper}.

\bibliography{paper}

\end{document}